\documentclass[referee]{gji}
\usepackage{timet}
\usepackage{graphicx}
\usepackage{lineno}

\title {High Altitude characterization of the Hunga Pressure Wave with Cosmic Rays by the HAWC Observatory}
 
\author[HAWC Collaboration]
  {
  \parbox{\linewidth}{\centering  
  R. Alfaro$^{1}$, C.~Alvarez$^{2}$,  J.C.~Arteaga-Vel\'azquez$^{3}$, K.P.~Arunbabu$^{4}$, D. Avila Rojas$^{1}$, H.A. Ayala Solares$^{5}$, R. Babu$^{6}$, E.~Belmont-Moreno$^{1}$, K.S.~Caballero-Mora$^{2}$, T.~Capistr\'an$^{7}$, A.~Carrami\~nana$^{8}$, S.~Casanova$^{9}$, O. Chaparro-Amaro$^{10}$, J.~Cotzomi$^{11}$, E.~De la Fuente$^{12}$, R.~Diaz Hernandez$^{8}$, M. A. DuVernois$^{13}$, K. Engel$^{14}$, C.~Espinoza$^{1}$, K.L.~Fan$^{14}$, N.~Fraija$^{7}$, J.A.~Garc\'ia-Gonz\'alez$^{15}$, F.~Garfias$^{7}$, M.M.~Gonz\'alez$^{7}$, J.A.~Goodman$^{14}$, S. Hernandez$^{1}$, D.~Huang$^{6}$, F.~Hueyotl-Zahuantitla$^{2}$, P.~H\"untemeyer$^{6}$, A.~Iriarte$^{7}$, V. Joshi$^{16}$, S. Kaufmann$^{17}$, A.~Lara$^{4}$\thanks{alara@igeofisica.unam.mx}, H. Le\'on Vargas$^{1}$\thanks{hleonvar@fisica.unam.mx}, J.T. Linnemann$^{18}$, A.L.~Longinotti$^{7}$, G.~Luis-Raya$^{17}$, K.~Malone$^{19}$, J.~Mart\'inez-Castro$^{10}$, J.A.~Matthews$^{20}$, P.~Miranda-Romagnoli$^{21}$, J.A.~Morales-Soto$^{3}$, E.~Moreno$^{11}$, L.~Nellen$^{22}$, R.~Noriega-Papaqui$^{21}$, N.~Omodei$^{23}$, E.G.~P\'erez-P\'erez$^{17}$, C.D.~Rho$^{24}$, D.~Rosa-Gonz\'alez$^{8}$, E. Ruiz-Velasco$^{25}$, H. Salazar$^{11}$, A. Sandoval$^{1}$\thanks{asandoval@fisica.unam.mx}, M. Schneider$^{14}$, A.J. Smith$^{14}$, Y. Son$^{24}$, O. Tibolla$^{17}$, K.~Tollefson$^{18}$, I.~Torres$^{8}$, R.~Torres-Escobedo$^{26}$, R. Turner$^{6}$, F.~Ure\~na-Mena$^{8}$, E. Varela$^{11}$, L.~Villase\~nor$^{11}$, X. Wang$^{6}$, E. Willox$^{14}$, H.~Zhou$^{26}$, C.~de Le\'on$^{3}$}\\ 
  $^{1}$ Instituto de F\'isica, Universidad Nacional Aut\'onoma de M\'exico, Ciudad de M\'exico, M\'exico \\
  $^{2}$ Universidad Aut\'onoma de Chiapas, Tuxtla Guti\'errez, Chiapas, M\'exico \\
  $^{3}$ Universidad Michoacana de San Nicol\'as de Hidalgo, Morelia, M\'exico \\
  $^{4}$ Instituto de Geof\'{i}sica, Universidad Nacional Aut\'onoma de M\'exico, Ciudad de Mexico, Mexico \\
  $^{5}$  Department of Physics, Pennsylvania State University, University Park, PA, USA\\
  $^{6}$  Department of Physics, Michigan Technological University, Houghton, MI, USA\\
  $^{7}$  Instituto de Astronom\'{i}a, Universidad Nacional Aut\'onoma de M\'exico, Ciudad de Mexico, Mexico\\
  $^{8}$ Instituto Nacional de Astrof\'{i}sica, \'Optica y Electr\'onica, Puebla, Mexico \\
  $^{9}$  Institute of Nuclear Physics Polish Academy of Sciences, PL-31342 IFJ-PAN, Krakow, Poland\\
  $^{10}$ Centro de Investigaci\'on en Computaci\'on, Instituto Polit\'ecnico Nacional, M\'exico City, M\'exico \\
  $^{11}$ Facultad de Ciencias F\'{i}sico Matem\'aticas, Benem\'erita Universidad Aut\'onoma de Puebla, Puebla, Mexico \\
  $^{12}$ Departamento de F\'{i}sica, Centro Universitario de Ciencias Exactas e Ingenierias, Universidad de Guadalajara,\\  Guadalajara, Mexico \\
  $^{13}$ Department of Physics, University of Wisconsin-Madison, Madison, WI, USA \\
  $^{14}$ Department of Physics, University of Maryland, College Park, MD, USA \\
  $^{15}$ Tecnologico de Monterrey, Escuela de Ingenier\'{i}a y Ciencias, Ave. Eugenio Garza Sada 2501, \\ Monterrey, N.L., Mexico, 64849 \\
  $^{16}$ Erlangen Centre for Astroparticle Physics, Friedrich-Alexander-Universit\"at Erlangen-N\"urnberg, Erlangen, Germany \\
  $^{17}$ Universidad Politecnica de Pachuca, Pachuca, Hgo, Mexico \\ 
  $^{18}$ Department of Physics and Astronomy, Michigan State University, East Lansing, MI, USA \\
  $^{19}$ Physics Division, Los Alamos National Laboratory, Los Alamos, NM, USA \\
  $^{20}$ Dept of Physics and Astronomy, University of New Mexico, Albuquerque, NM, USA \\
  $^{21}$ Universidad Aut\'onoma del Estado de Hidalgo, Pachuca, Mexico \\
  $^{22}$ Instituto de Ciencias Nucleares, Universidad Nacional Aut\'onoma de Mexico, Ciudad de Mexico, Mexico \\
  $^{23}$ Department of Physics, Stanford University: Stanford, CA 94305–4060, USA \\
  $^{24}$ University of Seoul, Seoul, Rep. of Korea \\  
 $^{25}$  Max-Planck Institute for Nuclear Physics, 69117 Heidelberg, Germany \\	
 $^{26}$ Tsung-Dao Lee Institute \& School of Physics and Astronomy, Shanghai Jiao Tong University, \\ Shanghai, People’s Republic of China   
 }

\date{   }
\pagerange{\pageref{firstpage}--\pageref{lastpage}}
\volume{X}
\pubyear{2022}

\begin{document}

\label{firstpage}

\maketitle


\begin{summary}
High-energy cosmic rays that hit the Earth can be used to study large-scale atmospheric perturbations. After a first interaction in the upper parts of the atmosphere, cosmic rays produce a shower of particles that sample the atmosphere down to the detector level. The HAWC (High-Altitude Water Cherenkov) cosmic-ray observatory in Central Mexico at 4,100 m elevation detects air shower particles continuously with 300 water Cherenkov detectors with an active area of 12,500 m$^{2}$. On January 15th, 2022, HAWC detected the passage of the pressure wave created by the explosion of the Hunga volcano in the Tonga islands, 9,000 km away, as an anomaly in the measured rate of shower particles. The HAWC measurements are used to characterize the shape of four pressure wave passages, determine the propagation speed of each one, and correlate the variations of the shower particle rates with the barometric pressure changes, extracting a barometric parameter. The profile of the shower particle rate and atmospheric pressure variations for the first transit of the pressure wave at HAWC is compared to the pressure measurements at Tonga island, near the volcanic explosion. This work opens the possibility of using large particle cosmic-ray air shower detectors to trace large atmospheric transient waves.
\end{summary}


\section{Introduction}

Large area cosmic-ray detectors at high elevations are a valuable tool for studying large perturbations of the Earth's atmosphere \cite{ab1}\cite{ab2}. The cosmic rays, predominantly protons, arrive uniformly from all directions and have their first interaction at the upper layers of the atmosphere, mainly with nitrogen, oxygen or argon nuclei at typical heights of 15 to 35 km. They initiate a cascade of secondary particles that propagates through the atmosphere at almost the speed of light, reaching the maximum number of produced secondary particles at an altitude of about 6 km \cite{ac1} \cite{ac2}. Beyond this shower maximum, the number of particles diminishes as they start to be absorbed or decay, until they reach the Earth surface. During this process, the shower particles sample a large volume of the atmosphere from the height of the first interaction point down to the ground level. Given the high flux of primary cosmic rays (AMS-2 measured 1,600 protons m$^{-2}$s$^{-1}$ with energies above 10 GeV \cite{bosc}\cite{aguilar}), a large cosmic-ray detector at a high elevation, near the shower maximum, can detect changes in the atmospheric mass density (pressure and temperature) to better than one part in 6,000 in the sampled volume \cite{grb}.

On January 15th, 2022, the Hunga volcano, in several explosive events, created atmospheric pressure waves that propagated around the globe several times \cite{ac1b} \cite{ac2b} \cite{ac3b}. This event is by far the largest volcanic explosion in modern times. It destroyed most of the Hunga Tonga and Hunga Ha`apai islands as well as the land bridge and the crater that had formed between them in previous eruptions \cite{ad1}. The eruption ejected tephra and gas up to the troposphere and produced waves that propagated through the Earth, the oceans, the atmosphere, and the ionosphere \cite{ad1} \cite{ad2}.

The HAWC Gamma-Ray observatory \cite{a} \cite{b}, in continuous operation since March 2015, is situated on the slopes of a stratovolcano in central Mexico at 4,100 m elevation. An undisturbed path over the Pacific Ocean connects the Hunga volcano to HAWC. Therefore, HAWC is well placed to characterize the shape of the Lamb atmospheric pressure waves created during the explosion. 

\section{The HAWC Cosmic-Ray Detector}

\begin{figure}
\centering
\includegraphics[width=8.6cm]{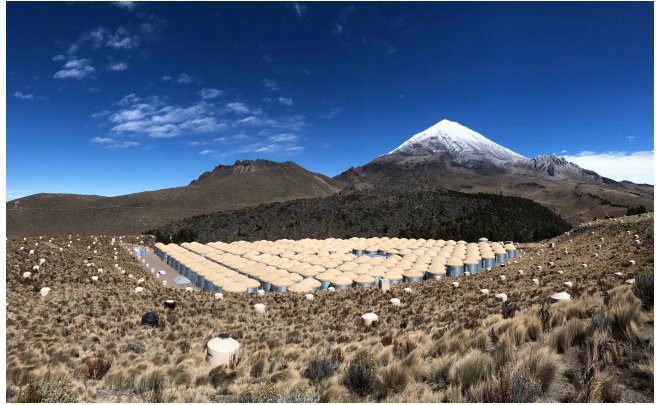}%
\caption{The HAWC gamma- and cosmic-ray observatory on the slopes of the Sierra Negra Volcano, Puebla, Mexico, at 4,100 m elevation. Air shower particles are detected in the 300 water tanks by the Cherenkov light they emit when passing through the water. The picture also shows the detector upgrade that consists of smaller WCDs that surround the main array.\label{fig:HAWCObs}}
\end{figure}

\begin{figure}
\centering
\includegraphics[width=5cm]{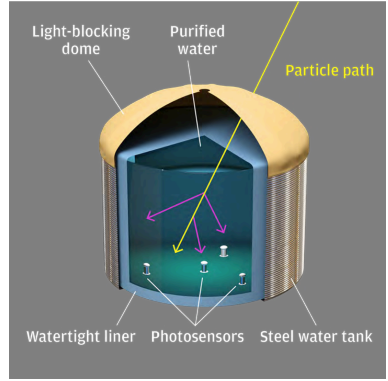}%
\caption{Sketch of one of the 300 water Cherenkov detectors of the HAWC observatory. They consist of a corrugated steel cylinder of 7.3 m diameter and 4.5 m height containing a specially made airtight and lighttight bag with 180,000 liters of highly purified water. The four photosensors anchored at the bottom only see the flashes of blue light created by relativistic shower particles as they traverse the water.\label{fig:HAWCWCD}}
\end{figure}

The High Altitude Water Cherenkov (HAWC) observatory is located on the slopes of the Sierra Negra volcano in the Pico de Orizaba National Park (latitude 18$\degr$59$\arcmin$41$\arcsec$ N, longitude 97$\degr$18$\arcmin$30.6$\arcsec$ W). In order to perform high-energy gamma-ray astronomy (from few hundred GeV up to few hundred TeV), it detects arriving air shower particles from the cascades initiated by primary gamma rays from space as they interact with the upper atmosphere. HAWC also registers the more numerous air shower events generated by primary cosmic rays that collide with the Earth's atmosphere. The observatory consists of an array of 300 water Cherenkov detectors (WCD), Fig. \ref{fig:HAWCObs}. Each WCD is a water tank of 7.3 m in diameter and 4.5 m in height, with a custom-made hermetic bag that contains 180,000 liters of ultra-pure water. The shower particles: electrons, positrons, pions, muons, and low energy gammas (which produce electrons via Compton scattering) entering the water volume at almost the speed of light produce Cherenkov light, flashes of blue light, that are detected by four very sensitive photomultiplier detectors (PMT's) anchored at the bottom of each of tank. The central one is a high quantum efficiency 10-inch Hamamatsu R7081-02 PMT.  Three 8-inch Hamamatsu R5912 PMTs are placed on the vertices of an equilateral triangle of 3.2 m sides, Fig. \ref{fig:HAWCWCD}. The array of 300 tanks spreads over 22,000 m$^{2}$ with an active area of 12,500 m$^{2}$ \cite{bc1}. 


Relativistic shower particles that enter the water volume produce Cherenkov light, and it is enough that one photoelectron is detected by the photosensors for the data acquisition system to record the event. In this way, all signals in the photomultipliers are digitized, recording their arrival time and amplitude. The data acquisition system also record the count rate of signals in each PMT in time intervals of 25 milliseconds and stores them in a file. HAWC aims to study cosmic rays and the sources of high-energy gamma rays, so the data acquisition system inspects all the arriving information and selects with an online computing farm only events from large showers having more than 28 signals in coincidence. The cosmic rays produce in HAWC a  trigger rate of 25 kHz, these events are recorded to disk producing 2 TByte of data daily. 

\section{The Observations}

For this analysis, in order to have the best accuracy in the determination of the rates, a selection of the photomultipliers was done. For the peripheral and central photomultipliers separately, a cut was made in the distribution of the standard errors calculated from the average rate of each PMT during the 4 days of this study, removing the outliers.  In total 927 photosensors were used, 215 of the central and 712 of the peripheral photomultipliers.  Variations of the count rate are due to mass density changes in the atmosphere (due to pressure and temperature variations), electric storms, space weather, and solar modulation of the cosmic-ray flux \cite{ab1}.     

A VAISALA PTB210 digital barometer, located inside a sea container to protect it from wind gusts, records the readings every second and it is synchronized in time with the cosmic-ray detector by the same GPS signal.  In this study, both the barometer data and the shower particle rates are averaged in 1-minute time intervals.

The dependence of the shower particle rate on variations of atmospheric conditions has been studied in the past \cite{ab1} \cite{ab2} \cite{bf1} \cite{ab4}. The results show an anti-correlation between pressure and the measured particle rate. Since the detectors are below the shower maximum, an increase in pressure increases the mass above the detector, creating greater absorption of the shower particles, which reduces the particle rate at the ground level.

It is also found \cite{bf1} \cite{bf2} \cite{bf3} and confirmed by model calculations of the development of atmospheric showers with particle transport codes like CORSIKA \cite{cor} that there is a linear relationship between the fractional change of the particle rate to the changes in pressure:

\begin{equation} \label{eq:deltar}
\Delta R/ <R> = \alpha \Delta P + b,
\end{equation}

where the proportional constant $\alpha$ is known as the barometric coefficient. 

\begin{figure}
\centering
\includegraphics[width=8.6cm]{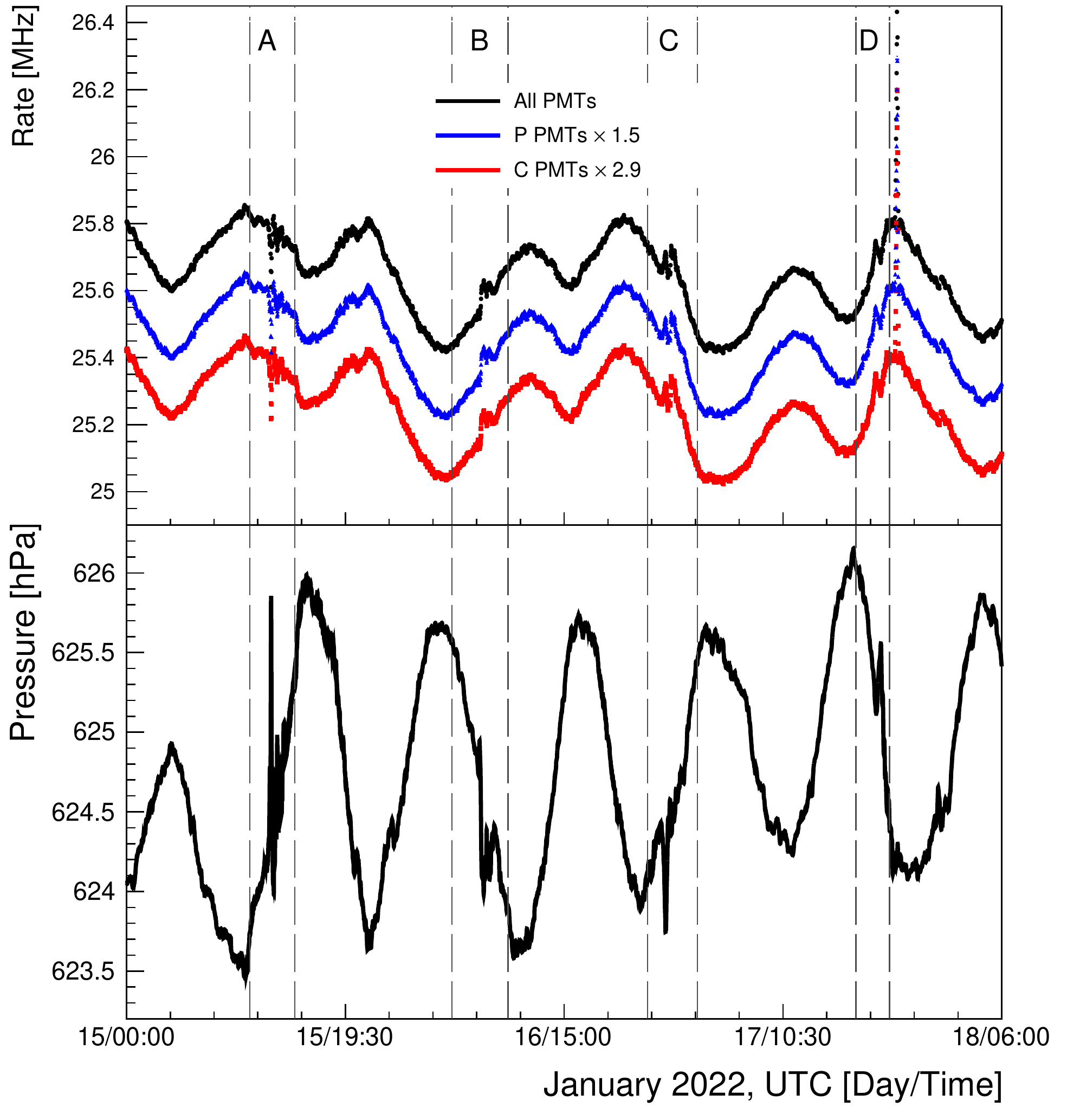}%
\caption{Top panel: Traces of the shower particle rates from the HAWC array as a function of time from January 15th, 00:00 UTC to January 18th, 06:00 UTC. Shown in black are the rates for all considered photosensors; in blue, the subset of peripheral detectors of each tank is scaled up by a 1.5 factor; in red, the subset of central detectors is scaled up by a 2.9 factor. The scale factors were used to display all the data with similar rate values. One sees the bi-diurnal oscillations and superimposed on them the signature of 4 passes of the Hunga explosion pressure wave delimited by the vertical lines in regions A-D. The spike after the D pass is due to an electric storm at the site. Bottom panel: Pressure variations from January 15th, 2022, 0:00 UTC to January 18th, 2022, 6:00 UTC. One sees the twice-daily oscillations are anti-correlated to the particle rate, and superimposed on them the pressure changes due to the passage of the pressure wave (A-D). The short path wave passages are (A) and (C), and the long path passages are (B) and (D).\label{fig:Traces}}
\end{figure}

The atmospheric pressure and the shower particle rate at the HAWC site shows a periodic variation every 12 hours because of its near-equatorial location at 19$\degr$N. These oscillations, well known for low-latitude regions, are due to atmospheric tides caused by the solar heating of the ozone and water vapor layers \cite{carr}. The maxima for the pressure occur at 10:00 and 22:00 hours local time. The shower particle count rate shows the same oscillations but it is anti-correlated with the pressure variations with maxima at 4:00 and 16:00 hours (Fig. \ref{fig:Traces}). One can see that during the data acquisition period, from  January 15 to 18th, 2022, on these semi-diurnal variations of the particle count rate and the atmospheric pressure, there are short-term disturbances superimposed ($\sim$2 hours duration) due to the passage of the pressure wave created by the explosion of the Hunga volcano, that took place just after $\sim$ 04:00 UTC on January 15 \cite{ac2b}. The change of  the count rate is shown in the top panel of Fig. \ref{fig:Traces} for the sum of all photosensors (black) and for the two subsets of peripheral (blue) and central (red) photomultipliers.

One can see four passages of the pressure wave (labelled in Fig. \ref{fig:Traces} as A-D). The first (A) corresponds to the pressure wave travelling the short arm of the great circle joining Hunga with the HAWC site through the Pacific Ocean. The second (B) corresponds to the wave travelling in the opposite direction along the great circle, passing through the antipodal point in Africa and arriving at HAWC from the Atlantic side. The third (C) is the (A) wave that, after passing over HAWC, continued to the antipodal point, then to Hunga, and arrived again at the HAWC site from the Pacific direction. The fourth (D) is the (B) wave that continued to Hunga and then to the antipodal point arriving at HAWC from the Atlantic. Along the short path, 95\% is over water, and entering the Mexican State of Guerrero, the highest mountains are in the Oriental Volcanic Rift in Mexico with 2,200 m elevation. Along the long path, 50 \% is over water, and on land, the highest elevations are on the Ethiopian High Lands, with peaks of 2,600 m elevation. The atmospheric pressure variations for these days show the same four passes of the pressure wave as the shower particle rate measurements, bottom panel of Fig. \ref{fig:Traces}.

\begin{figure}
\centering
\includegraphics[width=8.6cm]{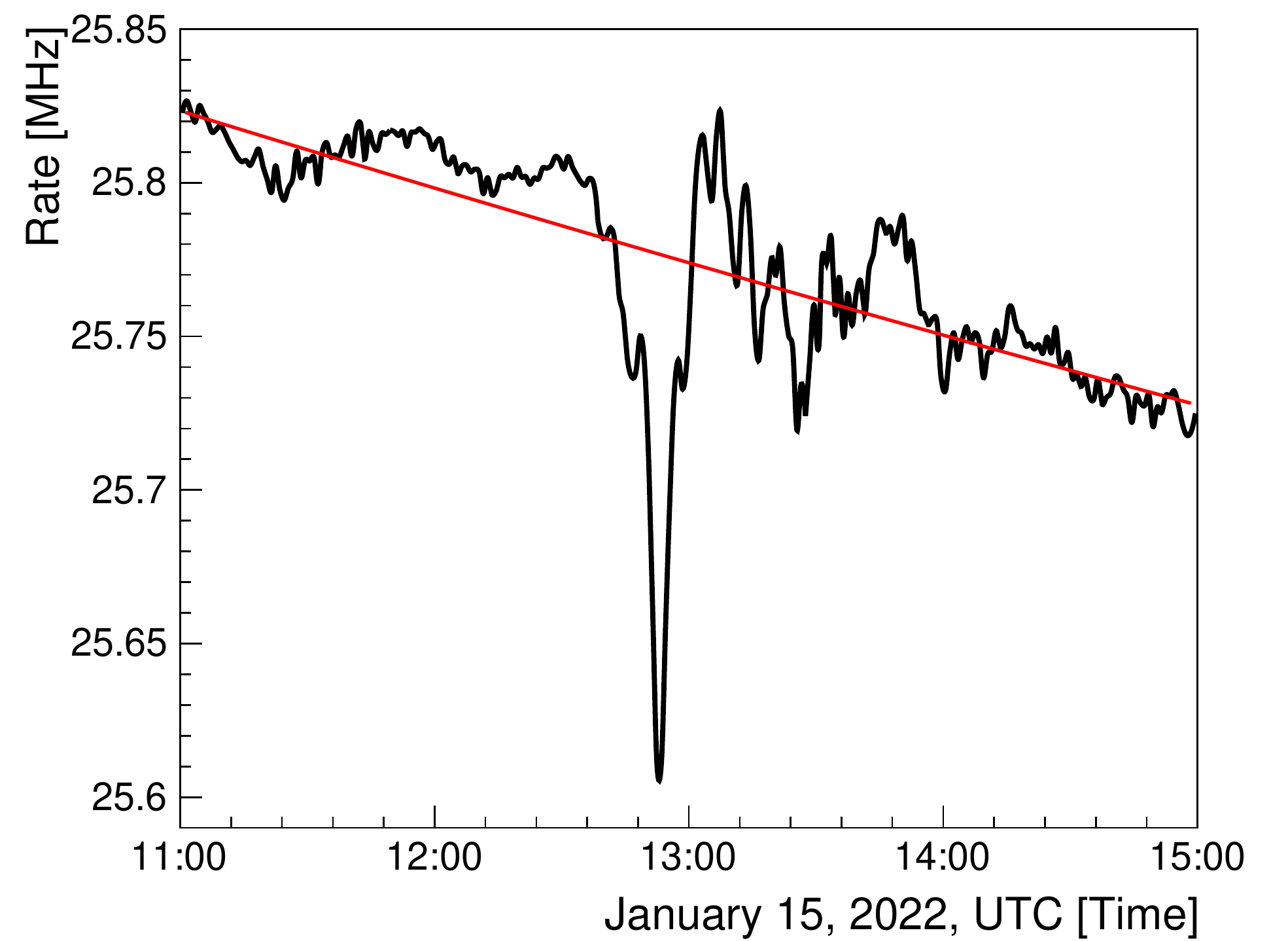}%
\caption{Expanded view of the first pass (A) of the Hunga pressure wave as recorded by the change of the rate (as measured by the sum of all PMTs) of detected shower particles from the HAWC array. Shown by a red line is the second order polynomial fitted to the data 2 hours before the wave arrival and two hours after the wave departure that is used to subtract the effect of the daily oscillations on which the pressure wave signal is superimposed. \label{fig:RateP1}}
\end{figure}

\begin{figure}
\centering
\includegraphics[width=8.6cm]{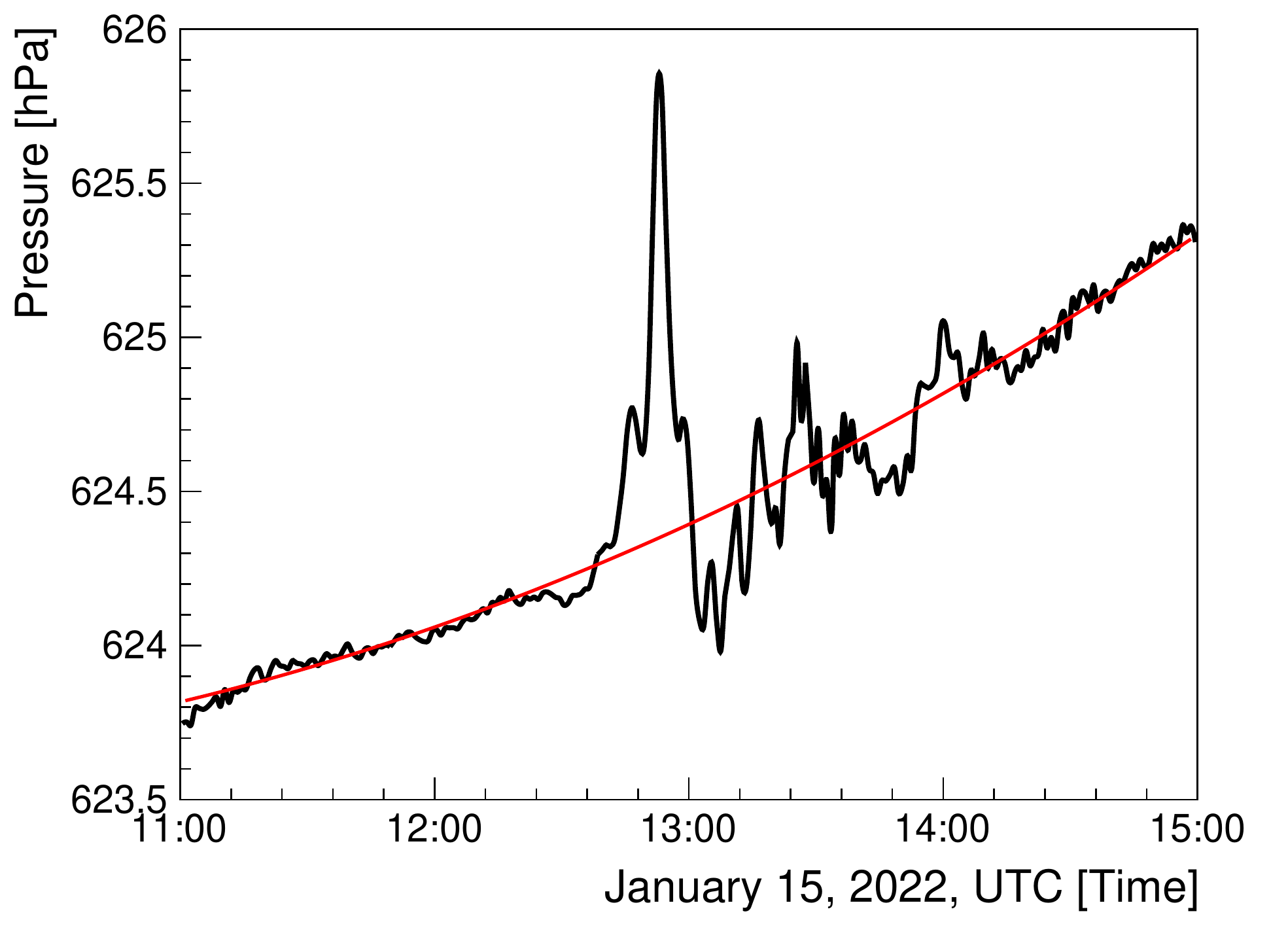}%
\caption{Pressure variations during the first pass (A) of the Hunga pressure wave from January 15th, 2022, 11:00 to 15;00 UTC, as recorded by the barometer at HAWC. A second order polynomial is fitted to the data from 2 hours before the pressure wave arrival and 2 hours after the wave departure, red line. The twice-daily oscillation effect on the data is removed by subtracting the fitted polynomial values. \label{fig:PressP1}}
\end{figure}

In order to extract the count rate and pressure variations due to the Hunga pressure wave, the effects of the semi-diurnal oscillations must be subtracted. Since the Hunga signal for each pass extends only for 2 hours, a simple procedure is applied. The count rate and pressure variations two hours before the wave's arrival and two hours after the wave departure are fitted by a second-degree polynomial and subtracted as an average value. The fit was done with $\chi^{2}$ minimization taking into account the uncertainties in the data points, which are the standard error in the mean along each axis. Fig. \ref{fig:RateP1} and Fig. \ref{fig:PressP1} show the variations of the rate and pressure for January 15th, 2022, from 11:00 to 15:00 UTC. The red line gives the second-degree polynomial fit. 

\begin{figure}
\centering
\includegraphics[width=8.6cm]{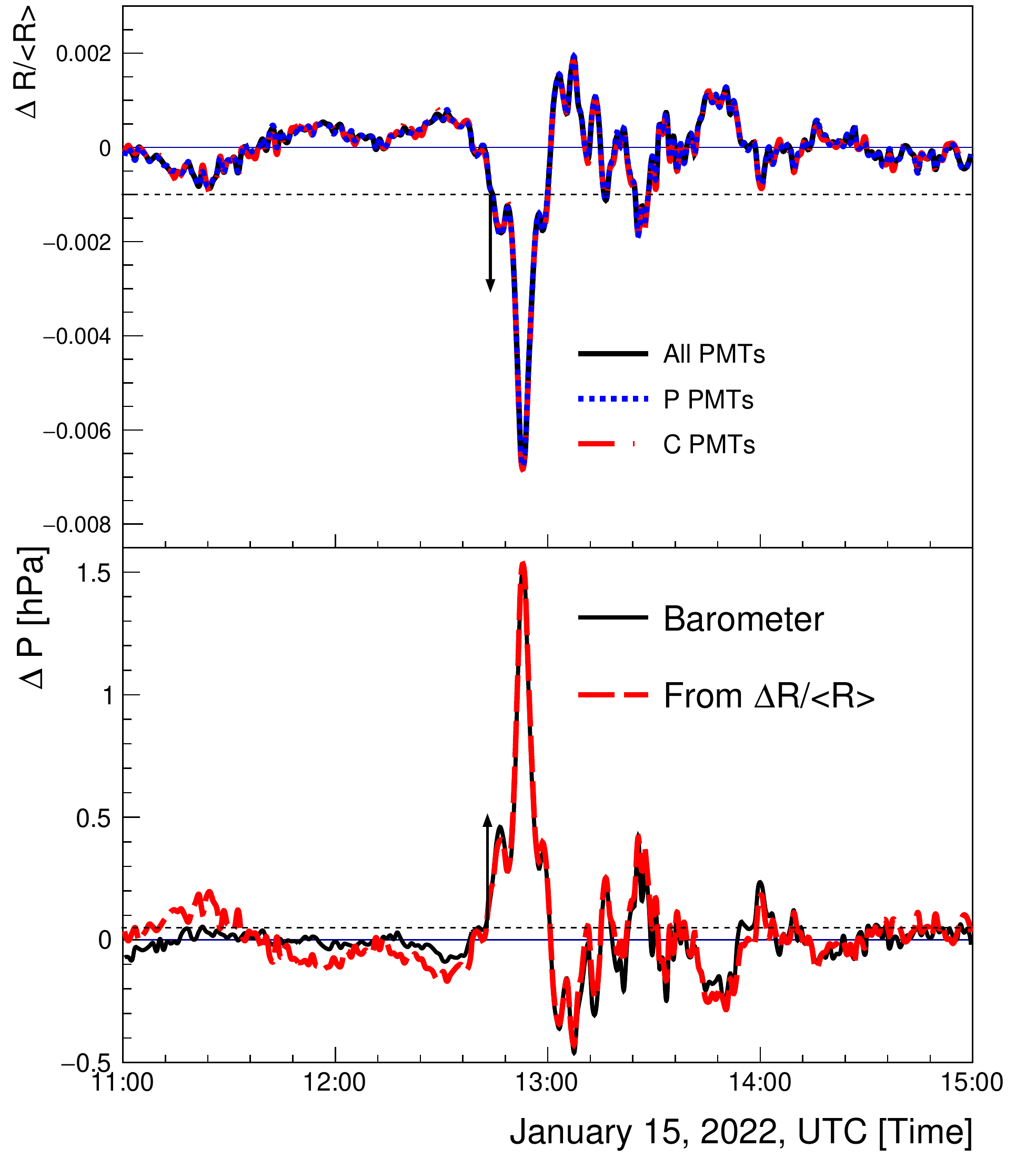}%
\caption{Top panel: Variations on the fractional rate of detected shower particles after the effect of the daily oscillation has been subtracted for the time of the first passage of the Hunga pressure wave. The black line shows the data for all photosensors, and the red and blue lines correspond to each tank's two subsets of central and peripheral sensors. The horizontal dashed line marks the fractional rate of the most significant fluctuation before the pressure wave's arrival. The point where the signal crosses this line defines the arrival time of the wave. It is indicated by the location of the arrow. Bottom panel: The black line shows the measured barometric pressure variations, with the arrival of the pressure wave indicated with the vertical arrow. The pressure variations calculated from the HAWC measured rate variations (top panel) using Eq. \ref{eq:deltar} is shown by the red line. \label{fig:Fluctuations}}
\end{figure}

A more sophisticated procedure is described in \cite{ab1} \cite{ak} to subtract the periodic pressure oscillations for periods of weeks to months to study heliospheric phenomena. Since the pressure wave passage is a short term anomaly compared to the diurnal oscillations, the simpler approach is used to conserve the fine detail features of this particular phenomenon.

The top panel of Fig. \ref{fig:Fluctuations} shows the resulting fractional change of the rate ($\Delta R/<R>$) for all photomultipliers as a black line and the two subsets of it: of the central PMTs only, shown by the red dashed line, and for the peripheral PMTs as the blue dashed line, for which the same procedure was applied. That the two subsets coincide exactly with each other shows that the procedure is consistent.

The bottom panel of Fig. \ref{fig:Fluctuations} shows the atmospheric pressure variations ($\Delta P$) for the same period of time as a black line.  For HAWC, the barometric coefficient in Eq. \ref{eq:deltar} was determined for a period of several months \cite{ab1} to be $\alpha$ = $-0.3426 \pm 0.0004$. Utilizing this value one can obtain from the measured differential rate variations (top panel of Fig. \ref{fig:Fluctuations}) the associated pressure variations (red curve in the bottom panel of Fig. \ref{fig:Fluctuations}). The very good agreement with the measured barometric pressure demonstrates that the determination by HAWC of the rate of cosmic shower particles makes it into a barometer with an active area of 12,500 m$^2$, sampling the atmospheric mass density in a cone of $\pm$ 50$\degr$ above it every 25 ms.

\begin{table}
  \begin{center}
    \caption{Characteristics of the structures around the main peak in the rate and pressure fluctuations data. $\Delta t$ is the time difference of the arrival times of lobes with respect to that of the central peak. It is computed for the rate as well as for the pressure signals.}
    \label{tab:table1}
    \begin{tabular}{cccccc}
		\hline	\hline
      Site & Feature & $\Delta$t for rate  & $\Delta$R/$<$R$>$  & $\Delta$t for P & $\Delta$P  \\
         &   & [min]  &  & [min] & [hPa] \\
		\hline
      HAWC & First lobe & -7 & -0.0018 & -7 & +0.45 \\       
       		\hline
      HAWC & Peak & 0 & -0.0066 & 0 & +1.5 \\
      		\hline
      HAWC & Second lobe & +6 & -0.0018 & +6 & +0.35 \\
      		\hline
      Tonga & First lobe & - & - & -5 & -10 \\
      		\hline	
      Tonga & Peak & - & - & 0 & -19 \\
            \hline
      Tonga & Second lobe & - & - & +8 & -10 \\       
      	\hline \hline
    \end{tabular}
  \end{center}
\end{table}

\begin{figure}
\centering
\includegraphics[width=8.6cm]{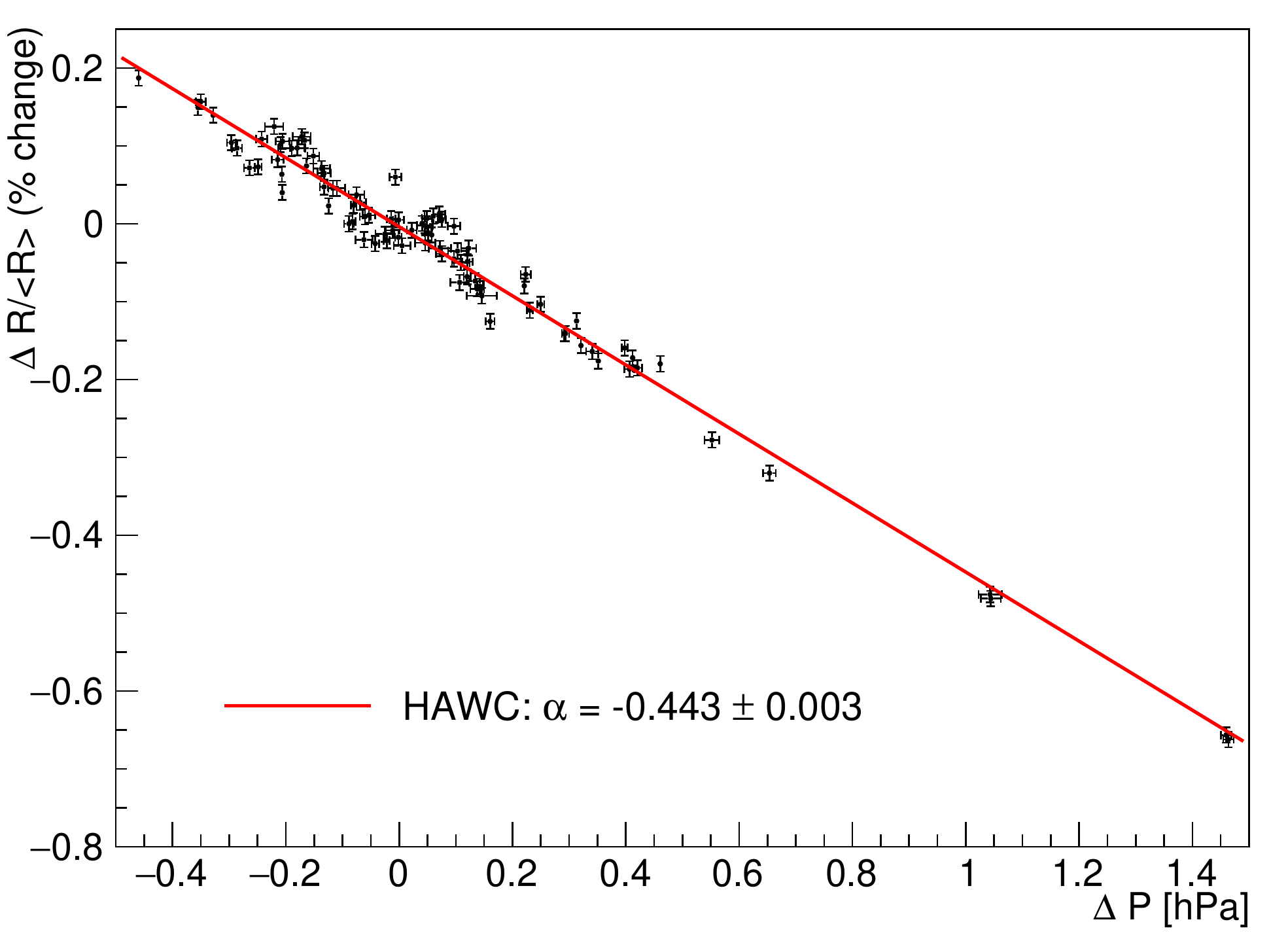}%
\caption{Correlation diagram of the percental rate change and the pressure variation for synchronous measurements during the passage of the first pressure wave from the Hunga explosion as recorded by the HAWC observatory in the time period from 12:40 to 14:10 UTC on January 15th, 2022. The points show the standard error on the mean for $\Delta P$ and $\Delta R/<R>$.  \label{fig:Correlation}}
\end{figure}

It is of interest to see what is the rate-pressure correlation for the short time span during the first passage of the pressure wave. Fig. \ref{fig:Correlation} shows the correlation between the measured percentile change of the shower particle rates integrated over 1 minute with the instantaneous barometric pressure change for the one and a half hours after 12:40 (UTC) on January 15th 2022 during the first passage. The fitted linear correlation gives a barometric coefficient of $\alpha$ = -0.433 $\pm$ 0.003 with a correlation coefficient of -0.988.    

One sees that the shape of the pressure wave at the HAWC site at 4,100 m elevation does not correspond to the shape of a classical Lamb wave \cite{lamb} \cite{amores}  that are detected at lower elevations all over the world, having a positive and negative peak of almost the same amplitude \cite{ac2b} \cite{ac3b}. Instead, one finds a high-pressure peak of 1.5 hPa flanked by two lobes of $\sim$1/3 amplitude, one at each side. The variations after the central peak show a complex structure that might have information about the eruption process; see below. An equivalent structure is seen in the measured rate fluctuations. Table \ref{tab:table1} gives the time difference between the lobes and the central peak and the amplitudes for the rate and pressure variations at the HAWC site.

The arrival time of the pressure wave is defined as the time when the signal exceeds the most significant fluctuation seen before the arrival time of the wave. This time is indicated by arrows in both panels of Fig. \ref{fig:Fluctuations}. The same analysis was done for the other three passes of the pressure wave in order to define their arrival times. The distance along the short arc from the volcano to the HAWC observatory was calculated using the output of the IDL function MAP$\_$2POINTS from the L3 Harris Geospatial software \cite{soft} and also using the Haversine formula \cite{hav} with an average value for the Earth radius of 6371 km.  The distance along the Long arc that passes through the antipode was calculated by subtracting the short arc distance from the Earth's circumference with an equatorial radius of 6378 km. The arrival times, reported in Table \ref{tab:table2}, are obtained with the threshold method using the rate and pressure data obtained at the HAWC site. The uncertainties reported in Table \ref{tab:table2} come from the difference in arrival times obtained using both data sets.

\begin{table}
  \begin{center}
    \caption{Arrival times and propagation speed of the different passes of the pressure wave as detected by HAWC at 4,100 m elevation in central Mexico.}
    \label{tab:table2}
    \begin{tabular}{ccccc}
		\hline	\hline
      Pass & Date [UTC] & Arrival time  [UTC Time]  & Speed [m/s]  & Type\\
		\hline
        A & Jan. 15 & 12:43 $\pm$ 00:01 & 316.2 $\pm$ 1.3 & Short  \\
		\hline
        B & Jan. 16 & 07:25 $\pm$ 00:01  & 312.4 $\pm$ 0.2  & Long \\	
        \hline				
        C & Jan. 16 & 23:48 $\pm$ 00:03  & 	324.9 $\pm$ 2.0  & Short \\
        \hline	
        D & Jan. 17 & 19:02 $\pm$ 00:05 & 	312.0 $\pm$ 1.3 & Long   \\

      	\hline \hline
    \end{tabular}
  \end{center}
\end{table}

The propagation speed of the pressure wave for the four passes as determined by the shower particle rate and the pressure variations are given in Table 2. One sees that the wave that initially travelled on the short arc is slightly faster than the one travelling on the long arc. The values agree with the determination by other authors \cite{ac2b} \cite{burt}.

\begin{figure}
\centering
\includegraphics[width=8.6cm]{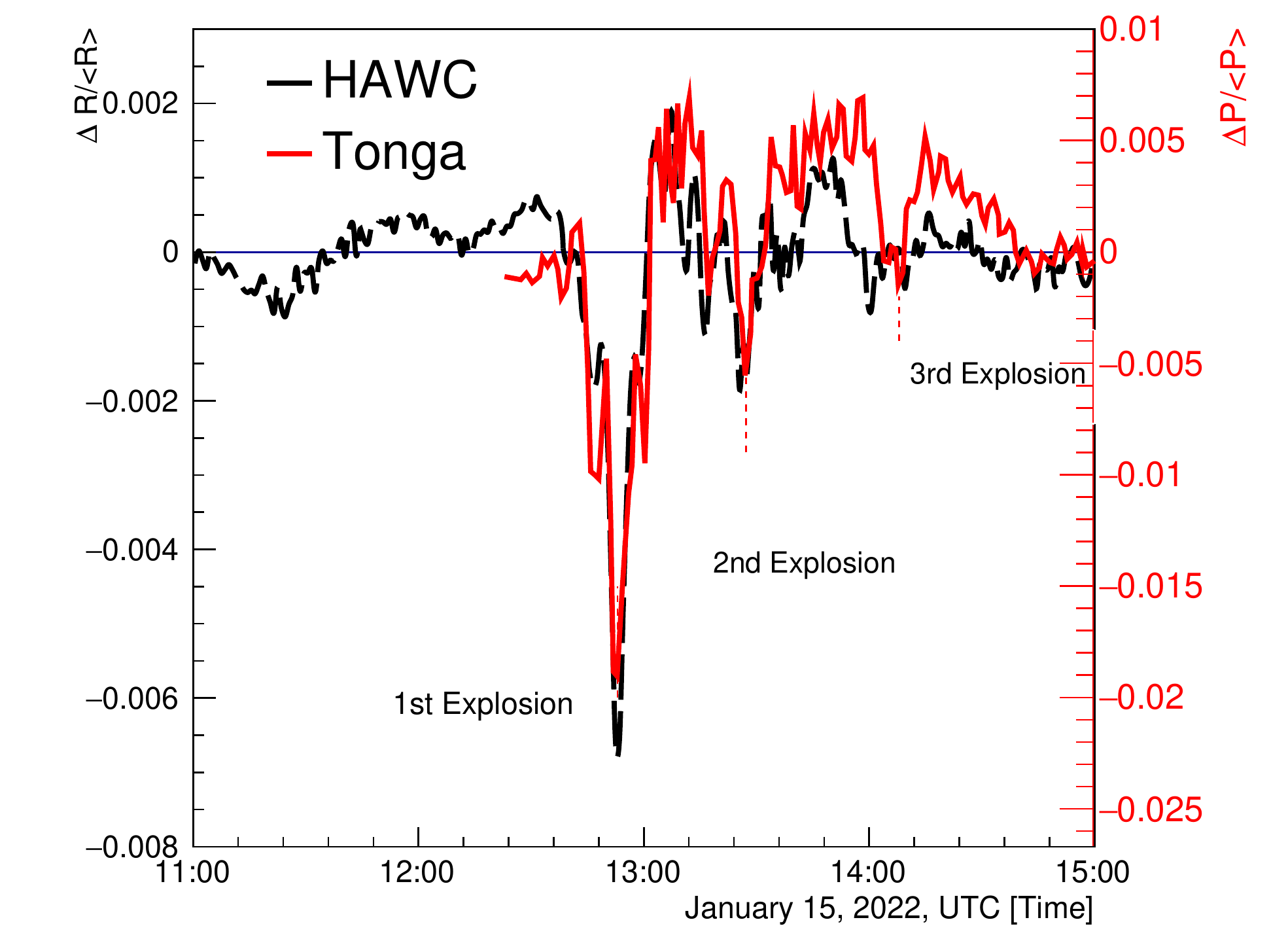}%
\caption{Comparison of two data sets for the first pass of the pressure wave, one from the barometric pressure recorded on the Tonga island, 64 km from the Hunga eruption (red) with the vertical scale on the right, and the other from the rate variations detected at the HAWC site 9000 km from Hunga (black). The pressure data from Tonga identifies three subsequent eruptions, from which the first one was the most violent. The pressure data from Tonga was shifted in time, so the most significant peaks coincide for comparison of the rate variations. \label{fig:ShapeComp}}
\end{figure}

There is a measurement of the barometric pressure close to the volcano on the Tonga island 64 km away, as reported by \cite{ac3b} in the Extended Data fig. 8. For this Tonga barometric data the effect of the daily pressure modulation was corrected by the same fitting procedure done to the HAWC pressure data. The resulting pressure anomalies are shown in Fig. \ref{fig:ShapeComp} with a red line and the vertical scale on the right. The same figure shows the HAWC fractional rate fluctuations for the first pass of the pressure wave. The Tonga data is shifted in time to have the main peaks coincide. In the Tonga data the authors identified four subsequent explosions, being the first one the most violent. For the first explosion both the Tonga and the HAWC data show a similar structure of a principal peak with two lower amplitude lobes on each side. Table \ref{tab:table1} summarizes the parameters of the two lobes and the prominent eruption peak for the Tonga pressure measurements. One sees that the amplitude of the principal peak is much greater than what is measured at HAWC, reaching 19 hPa. 

The pressure anomaly from the first pass of the pressure wave generated by the Hunga explosion shows an exponential attenuation as a function of the distance of the measurement ( \cite{ac3b}, Extended Data fig. 1) with

\begin{equation} \label{eq:PAt}
P = 15 e^{-r/2500},
\end{equation}

with P in hPa and r in km. The calculated amplitude at the HAWC site is 0.41 hPa, which is smaller than the observed value of 1.5 hPa, implying that the pressure wave is less attenuated at a high elevation. However, the fit quality is not very accurate at distances of 10,000 km, so it is not a solid conclusion.

The detailed study of the other transits of the Tonga pressure wave in the HAWC data is more complicated, but is in progress. The reason is that the wave's profile changes radically, and other effects, including changes in the atmosphere's temperature profile, moisture content, high electric fields or changes in the state of the heliosphere, must be evaluated.

\section{Conclusions}

The HAWC observatory was built to study the sources of the highest energy gamma rays in our galaxy and beyond \cite{3hwc}. These are the most violent regions in our Universe, like remnants of supernovas, accreting black holes, or active galaxies \cite{agn}. As part of the daily operations, it continuously records the rate of arrival of shower particles from all directions in the sky created by cosmic rays interacting in the upper atmosphere. Variations in the rate of arriving particles can be determined with a precision of 1 in 6,000. 

It is found that the anomalies of the rate of shower particles detected by the HAWC observatory at 4,100 m a.s.l. can be used to study large$-$scale atmospheric perturbations over a more extended region of the atmosphere than the vertical column sampled by barometers. This new method will complement other well$-$established procedures \cite{hang}.

During the January 15th$–$18th, 2022 period, the transits of the pressure wave created by the Hunga volcano explosion generated anomalies in the count rate of HAWC detector signals and pressure variations from an on site barometer. The arrival times of four passes of the wave were determined, and the resulting propagation velocities show that the portion of the Lamb wave traveling westward from Hunga is slightly faster than the part travelling to the East.  The anti-correlation of the rate and pressure variations during the passage of the first arrival is linear and surprisingly tight, with a barometric coefficient $\alpha$= -0.433 $\pm$ 0.003.

The anomaly from the first passage of the pressure wave has a complicated profile that does not correspond to the shape of a classical Lamb wave. The rate shows a deep minimum flanked from each side by two minima of one third amplitude and followed by a complicated structure. The anomalies after the main event agrees perfectly with the barometric pressure measurements from the Tonga island, 64 km from the volcanic eruption. They correspond to three of the four eruptions that occurred during the short period of an hour and a half.

Work is in progress on a detailed analysis of the other three passages of the soliton pressure wave created by the tremendous Hunga volcanic explosion.

\section{Data availability}

The HAWC total rate data will be shared on reasonable request to the authors. All figures with data have been produced using the ROOT analysis software \cite{root}. 

\begin{acknowledgments}
We acknowledge the support from: the US National Science Foundation (NSF); the US Department of Energy Office of High$-$Energy Physics; the Laboratory Directed Research and Development (LDRD) program of Los Alamos National Laboratory; Consejo Nacional de Ciencia y Tecnolog\'ia (CONACyT), M\'exico, grants 271051, 232656, 260378, 179588, 254964, 258865, 243290, 132197, A1$-$S$-$46288, A1$-$S$-$22784, c\'atedras 873, 1563, 341, 323, Red HAWC, M\'exico; DGAPA$-$UNAM grants IG101320, IN111716$-$3, IN111419, IA102019, IN110621, IN110521; VIEP$-$BUAP; PIFI 2012, 2013, PROFOCIE 2014, 2015; the University of Wisconsin Alumni Research Foundation; the Institute of Geophysics, Planetary Physics, and Signatures at Los Alamos National Laboratory; Polish Science Centre grant, DEC$-$2017/27/B/ST9/02272; Coordinaci\'on de la Investigaci\'on Cient\'ifica de la Universidad Michoacana; Royal Society $-$ Newton Advanced Fellowship 180385; Generalitat Valenciana, grant CIDEGENT/2018/034; The Program Management Unit for Human Resources \& Institutional Development, Research and Innovation, NXPO (grant number B16F630069); Coordinaci\'on General Acad\'emica e Innovaci\'on (CGAI$-$UdeG), PRODEP$-$SEP UDG$-$CA$-$499; Institute of Cosmic Ray Research (ICRR), University of Tokyo, H.F. acknowledges support by NASA under award number 80GSFC21M0002.  Thanks to Scott Delay, Luciano D\'iaz and Eduardo Murrieta for technical support.
\end{acknowledgments}

\label{lastpage}


\begin{thebibliography}{}

\bibitem[\protect\citename{Abeysekara et al.  }2015]{grb}
Abeysekara A.U. et al. (HAWC Collaboration) 2015, Search for gamma-rays from the unusually bright GRB 130427A with the HAWC gamma-ray observatory, \textit{The Astrophysical Journal}, \textbf{800}, 78.

\bibitem[\protect\citename{Abeysekara et al. }2017]{b}
Abeysekara , A.U. et al. (HAWC Collaboration) 2017. Observation of the Crab Nebula with the HAWC Gamma-Ray Observatory, \textit{The Astrophysical Journal }, \textbf{843}, 39.

\bibitem[\protect\citename{Abeysekara et al. }2018]{a}
Abeysekara , A.U. et al. (HAWC Collaboration) 2018. Data acquisition architecture and online processing system for the HAWC gamma-ray observatory, \textit{Nuclear Instruments and Methods in Physics Research Section A: Accelerators, Spectrometers, Detectors and Associated Equipment }, \textbf{888}, 138-146.

\bibitem[\protect\citename{Adam }2022]{ac1b}
Adam, D., 2022. Tonga volcano eruption created puzzling ripples in Earth’s atmosphere, \textit{Nature}, \textbf{601}, 497.

\bibitem[\protect\citename{M. Aguilar et al.  }2015]{aguilar}
 Aguilar M. et al. (AMS Collaboration) 2015, Precision Measurement of the Proton Flux in Primary Cosmic Rays from Rigidity 1 GV to 1.8 TV with the Alpha Magnetic Spectrometer on the International Space Station, \textit{Physical Review Letters}, \textbf{114}, 171103.

\bibitem[\protect\citename{Aharonian et al. }2022]{bd3}
Aharonian, F., et al. (LHAASO Collaboration) 2022. Flux Variations of Cosmic Ray Air Showers Detected by LHAASO-KM2A During a Thunderstorm on 10 June 2021, \textit{arXiv:2207.12601}.

\bibitem[\protect\citename{Akiyama et al. }2020]{ak}
  Akiyama, S., et al. (HAWC Collaboration) 2020. Interplanetary Magnetic Flux Rope Observed at Ground Level by HAWC, \textit{The Astrophysical Journal}, \textbf{905} 73.

\bibitem[\protect\citename{Albert et al.  }2020]{3hwc}
 Albert A., et al. (HAWC Collaboration), 2020. , 3HWC: The Third HAWC Catalog of Very-high-energy Gamma-Ray Sources, \textit{The Astrophysical Journal}, \textbf{905}, 76.

\bibitem[\protect\citename{Albert et al.  }2022]{agn}
 Albert A., et al. (HAWC Collaboration), 2022. , Long-term Spectra of the Blazars Mrk 421 and Mrk 501 at TeV Energies Seen by HAWC, \textit{The Astrophysical Journal}, \textbf{929}, 125.

 \bibitem[\protect\citename{Alfaro et al. }2017]{bc1}
  Alfaro, R., et al.,(HAWC Collaboration) 2017. All-particle cosmic ray energy spectrum measured by the HAWC experiment from 10 to 500 TeV., \textit{Physical Review D}, \textbf{96}, 122001.

\bibitem[\protect\citename{Alvarez et al. }2021]{ab1}
Alvarez C. et al., 2021. HAWC as a Ground-Based Space-Weather Observatory, \textit{Solar Physics}, \textbf{296}, 89.
    
 \bibitem[\protect\citename{Amores et al. }2022]{amores}
 Amores, A., et al., 2022. , Numerical Simulation of Atmospheric Lamb Waves Generated by the 2022 Hunga-Tonga Volcanic Eruption, \textit{Geophysical Research Letters}, \textbf{49}, 6.
    
\bibitem[\protect\citename{Arunbabu et al. }2017]{ab2}
 Arunbabu K.P. et al., 2017. Dependence of the muon intensity on the atmospheric temperature measured by the GRAPES-3 experiment, \textit{Astroparticle Physics}, \textbf{94}, 22-28.

\bibitem[\protect\citename{Bolognino }2012]{bd4}
 Bolognino, I., 2012. Study of the Influence of High Electric Field Variations on Cosmic Ray Flux detected by the ARGO-YBJ Experiment, \textit{Scientifica Acta}, \textbf{6}, 1.

\bibitem[\protect\citename{Boschini et al. }2014]{bosc}
Boschini M.J.,  Rancoita P.G. and Tacconi M., SR-NIEL — 7 \textit{Calculator: Screened Relativistic (SR) Treatment for Calculating the Displacement Damage and Nuclear Stopping Powers for Electrons, Protons, Light- and Heavy- Ions in Materials (version 9.0)};  [Online] available at INFN sez. Milano-Bicocca, Italy [2022, 09]:  http://www.sr-niel.org/.

\bibitem[\protect\citename{Brun \&  Rademakers }1996]{root}
Brun R. and Rademakers F., ROOT - An Object Oriented Data Analysis Framework,
\textit{Proceedings AIHENP'96 Workshop, Lausanne, Sep. 1996, Nuclear Instruments and Methods in Physics Research Section A: Accelerators, Spectrometers, Detectors and Associated Equipment 389 (1997) 81-86}. Release v6.18/04, 11/09/2019.

\bibitem[\protect\citename{Burt  }2022]{burt}
Burt S., 2022. , Multiple airwaves crossing Britain and Ireland following the eruption of Hunga Tonga-Hunga Ha'apai on 15 January 2022, \textit{Weather}, \textbf{77}, 3.

\bibitem[\protect\citename{Carrasco et al. }2009]{carr}
 Carrasco E., at al. 2009. , Weather at Sierra Negra: 7.3-yr statistics and a new method to estimate
the temporal fraction of cloud cover, \textit{Monthly Notices of the Royal Astronomical Society}, \textbf{398}, 407-421.

\bibitem[\protect\citename{De Mendonça  } 2013]{bf1}
 De Mendonça, R.R.S, et al. 2013. Analysis of atmospheric pressure and temperature effects
on cosmic ray measurements, \textit{Journal of Geophysical Research: Space Physics }, \textbf{118} 1403–1409.

\bibitem[\protect\citename{Engel et al. }2011]{ac1}
Engel, R., Heck, D., and  Pierog, T., 2011. Extensive Air Showers and Hadronic Interactions at High Energy, \textit{Annual Review of Nuclear and Particle Science}, \textbf{61}, 467-489.

\bibitem[\protect\citename{Heck et al. }1998]{cor}
 Heck D. et al, 1998. CORSIKA: a Monte Carlo code to simulate extensive air showers, \textit{Forschungszentrum Karlsruhe Report FZKA 6019}.

\bibitem[\protect\citename{Kampert \& Unger }2012]{ac2}
Kampert, K.H., and Unger, M., 2012. Measurements of the cosmic ray composition with air shower experiments, \textit{ Astroparticle Physics}, \textbf{35}, 660-678.

\bibitem[\protect\citename{Korn \& Korn  }2000]{hav}
Korn G.A., Korn T.M., 2000. Mathematical Handbook for Scientists and Engineers: Definitions, Theorems, and Formulas for Reference and Review \textit{Dover Publications, Inc.}.

\bibitem[\protect\citename{Lamb  }1932]{lamb}
 Lamb, H., 1932. , Hydrodynamics, \textit{Cambridge University Press}.

\bibitem[\protect\citename{Lin et al. }2022]{ad2}
 Lin, J.T., et al. 2022. Rapid Conjugate Appearance of the Giant Ionospheric Lamb
Wave Signatures in the Northern Hemisphere After Hunga-
Tonga Volcano Eruptions., \textit{Geophysical Research Letters}, \textbf{49}, 8.

\bibitem[\protect\citename{L3HarrisGeospatial  }2022]{soft}
 L3 Harris Geospatial, 2022. , \textit{L3Harris Geospatial} https://www.l3harrisgeospatial.com.

\bibitem[\protect\citename{Martinelle }1968]{ab4}
 Martinelle, S., 1968. Air pressure dependence of cosmic ray intensity, \textit{Tellus }, \textbf{20} 179-197.

\bibitem[\protect\citename{Matoza et al. }2022]{ac2b}
Matoza, R.S., et al., 2022. Atmospheric waves and global seismoacoustic observations of the January 2022 Hunga eruption, Tonga , \textit{Science}, \textbf{377}, 95-100.

\bibitem[\protect\citename{Otsuka }2022]{ad1}
Otsuka, S., 2022. Visualizing Lamb Waves From a Volcanic Eruption Using Meteorological Satellite Himawari-8., \textit{Geophysical Research Letters}, \textbf{49}, 8.

\bibitem[\protect\citename{Santos et al. } 2021]{bf2}
 Santos, N. A., et al. 2021. Observations of the cosmic ray detector at the Argentine
Marambio base in the Antarctic Peninsula, \textit{Proceedings of Science }, \textbf{ICRC2021} 304.

\bibitem[\protect\citename{Savi\'c et al.} 2021]{bf3}
Savi\'c, M., et al., 2021. New empirical methods for correction of meteorological
effects on cosmic ray muons, \textit{Proceedings of Science }, \textbf{ICRC2021} 1252.

\bibitem[\protect\citename{Su et al.  }2021]{hang}
Su, H., Yang, T., Sun, B., Yang, X., 2021. Modified atmospheric pressure extrapolation model using ERA5 for geodetic applications, \textit{GPS Solutions}, \textbf{25}, 118.

\bibitem[\protect\citename{Wright et al. }2022]{ac3b}
Wright, C., et al., 2022. Tonga eruption triggered waves propagating globally from surface to edge of space, \textit{Earth and Space Science Open Archive}, 2022.

\end{thebibliography}
\end{document}